\documentclass[aps,prx,reprint,twocolumn,
               amsmath,amssymb,amsfonts,
               notitlepage]{revtex4-1}
               
\AtBeginDocument{\usepackage{booktabs}}               
\makeatletter
\g@addto@macro\bfseries{\boldmath}
\makeatother

\usepackage{tikz}
\usetikzlibrary{arrows.meta, calc}

\usepackage[varg]{txfonts}
\usepackage[T1]{fontenc}
\usepackage[utf8]{inputenc}

\usepackage[varg]{txfonts}
\usepackage{hyperref}
\usepackage{amsmath}
\usepackage{color}
\usepackage{graphicx}
\usepackage[percent]{overpic}
\usepackage{mathrsfs}
\usepackage{bm}
\usepackage{braket}

\definecolor{cred}{RGB}{228,26,28}
\definecolor{cblue}{RGB}{8,48,107}
\definecolor{cgreen}{RGB}{77,175,74}
\definecolor{cgray}{RGB}{150,150,150}
\definecolor{clgray}{RGB}{200,200,200}
\definecolor{cpurple}{RGB}{152,78,163}
\definecolor{corange}{RGB}{255,127,0}
\definecolor{cgold}{RGB}{230,171,2}

\newcommand{\avg}[1]{\langle #1 \rangle}

\newcommand{\grad}{\vec{\nabla}}
\newcommand{\del} {\partial}

\renewcommand{\vec}[1]{\boldsymbol{#1}}
\newcommand{\mat}[1]{\vec{#1}}
\newcommand{\trp}[1]{{#1}^{\intercal}}
\newcommand{\vhat}[1]{\vec{\hat{#1}}}

\hypersetup{colorlinks=true,linkcolor=cblue,citecolor=cblue,urlcolor=cblue} 

\begin{document}

\title{Classical Kitaev model in a magnetic field}
\author{Paul A. McClarty}
\affiliation{Laboratoire Léon Brillouin, CEA, CNRS, Université Paris-Saclay, CEA Saclay, 91191 Gif-sur-Yvette, France.}
\author{Roderich Moessner}
\affiliation{Max-Planck-Institut f\"ur Physik komplexer Systeme, 01187 Dresden, Germany}
\author{Karlo Penc}
\affiliation{Institute for Solid State Physics and Optics, HUN-REN Wigner Research Centre for Physics, P.O.B. 49, H-1525 Budapest, Hungary}
\author{Jeffrey G. Rau}
\affiliation{Department of Physics, University of Windsor, 401 Sunset Avenue, Windsor, Ontario, N9B 3P4, Canada}

\begin{abstract}
Motivated by experiments on spin-orbit coupled magnets with Kitaev exchange in magnetic fields, we present an analysis of the classical Kitaev honeycomb model in the presence of a magnetic field. We show that there is a spin liquid regime that exists within a finite window of fields from zero up to a finite threshold before transitioning into the polarized paramagnet. We uncover constraints that spins need to satisfy in the ground state and show that they determine the exact limiting zero temperature behavior of the heat capacity and magnetic susceptibility within the spin liquid as a function of field. When the field is finite, both the two-point spin and the quadrupolar correlations are short-ranged, in contrast to the zero-field case. We rationalize an effective mass for the quadrupolar correlations in terms of a coarse-grained theory with fluctuating effective charge degrees of freedom. Finally, we show that weak site-dilution does not change
 the magnetization within the spin liquid -- a kind of ``perfect'' compensation of the site dilution.
\end{abstract}

\date{\today}

\maketitle


\section{Introduction}

Models with highly degenerate, locally constrained ground states are of considerable interest in condensed matter and statistical physics. Such models are foundational to frustrated magnetism and indeed a large class are geometrically frustrated antiferromagnets with Ising or vector spins \cite{lacroix2011introduction}. The low energy constrained degenerate manifold of states  map to a rich tapestry of other important classical models and the discrete models, at least, are often precursors to quantum liquids. Some of the importance attached to these models is tied to the fact that they typically have a coarse-grained description in terms of some continuum gauge theory thus providing relatively straightforward examples of fractionalization in a classical context. A notable example in three dimensions is the {\it Coulomb phase} in which the local constraints in the ground state of the lattice model can be captured by an emergent $U(1)$ gauge field and where the power law correlations in the ground state are inherited from an emergent electrostatics~\cite{Youngblood1981,Huse2003,Henley2005,henley2010coulomb}. Such Coulomb phases and their characteristic \emph{pinch point} correlations in momentum space \cite{Isakov2004} are relevant to materials such as water ice and the dipolar spin ice materials~\cite{bramwell2001spin,castelnovo2012spin,henley2010coulomb,udagawa2021spin} and, for example, NaCaNi$_2$F$_7$~\cite{zhang2019} that approximately realizes the pyrochlore Heisenberg antiferromagnet~\cite{moessner1998a} at intermediate temperatures.

While Coulomb phases have been recognized in some form for several decades, relatively recent progress has enlarged the nature and type of such classical correlated liquids beyond Coulomb phases. The cases can broadly be categorized into two classes. The first are correlated liquids with power law correlated spins such as higher rank spin liquids where the Gauss law of an emergent electrostatics is replaced by a multicomponent Gauss law acting on a rank-$2$ (or higher) tensor field~\cite{LH2018,Yan2020}. The second are comparatively featureless liquids with exponential correlations~\cite{Rehn2016,yanpohle2025}. An important tool for understanding and, more recently, cataloguing~\cite{davier2023,yan2024,yan2024b,fang2024} different classical spin liquids is to reduce the interacting spin model to a single particle band structure -- an approach that can be justified when the number of spin components is large. In such a classical spin liquid, the low-lying bands are completely flat encoding both the macroscopic degeneracy and the local constraint. Liquids with power law and exponential spin correlations are revealed through whether the flat bands are gapless or gapped respectively. A large class of flat band models can be constructed systematically.  While these constructions have proven useful in exploring the wealth of possible spin liquid phases that can appear in classical models -- the above-mentioned approach, perhaps surprisingly frequently, describes even Ising or Heisenberg spin models -- there are examples of classical spin liquids that fall outside this classification. 

An important example is the classical Kitaev honeycomb model. This model is known to have lowest energy states forming a classical spin liquid with an extensive ground state entropy determined by local constraints on the spins \cite{baskaran2008,chandra2010}. Crucially, the non-linear length constraint is essential and thus this phase does not neatly fall into the classification discussed above. Accordingly, while the spin-spin correlations are short-range -- zero beyond nearest neighbours -- pinch point correlations manifest in the quadrupolar correlations of the spins instead \cite{chandra2010}. So this classical liquid is a Coulomb phase in the \emph{quadrupolar} degrees of freedom -- a feature that is rare among classical spin liquids, particularly where the interactions are only between dipolar degrees of freedom \cite{pohle2026spinliquidsspin1pyrochlore}. 

Investigations into the classical model were motivated by its famous spin one-half counterpart that has a gapless {\it quantum} spin liquid ground state \cite{kitaev2006anyons} with emergent Majorana fermions excitations and gapped visons. Within the condensed matter community alone, the spin one-half model has inspired a large number of theoretical and experimental works with the larger goal of realizing the quantum spin liquid in materials or in quantum simulators~\cite{jackeli2009,Chaloupka2010,katukuri2014kitaev,rau2016spin,hermanns2017physics,trebst2017kitaev,winter2017,winter2016challenges}. In the solid state, $\alpha$-RuCl$_3$ is believed to have relatively strong Kitaev exchange couplings \cite{moller2025}. This material also has anomalously broadened magnetic excitations about the ordered low temperature magnetic phase. In an in-plane magnetic field, the magnetic order is suppressed, giving way to a new phase in which continuum inelastic scattering and a thermal Hall response have been reported \cite{Kasahara2018}. These experimental results have inspired theorists to consider the nature of the Kitaev model at finite field and in the presence of exchange couplings that are presumably present in the material alongside the Kitaev coupling \cite{janssen2016,janssen2019,yip2022,franke2022,f4b4-h1yr,holdhusen2024}. The magnetic field phase diagram of these models is extremely rich. Even in the pure Kitaev spin one-half model in finite field between the low field chiral spin liquid phase and the high field paramagnet, the model exhibits one or several intermediate phases whose nature remains controversial \cite{gohlke2018,zhu2018,wang2019,patel2018,gordon2019,hickey2019,zhang2022}. 

In this paper, we present results on the classical Kitaev honeycomb model in a magnetic field. We show that the finite field behavior of the {\it classical} vector spin antiferromagnetic Kitaev honeycomb model exhibits an intermediate field classical spin liquid phase of its own, that is distinct from the zero field classical spin liquid. Often classical spin liquids are fragile, in that natural physical perturbations tend to lift the extensive degeneracy. 
Remarkably, here an application of a magnetic field does not preserve the liquid nor does it destroy it entirely but instead changes its nature. We completely characterize the resulting phase using a combination of numerical simulations and analytical arguments. We show that the thermodynamics has the hallmarks of a classical spin liquid down to zero temperature without the presence of order by disorder. The spin correlations are short-ranged, even down to zero temperature, and remain so at finite field. The quadrupoles, which are known to have pinch point correlations in zero field and at zero temperature, acquire a finite width once the magnetic field is non-zero.  
We rationalize the width within the context of a coarse-grained theory where charge fluctuations introduce a mass gap into the liquid at finite field $-$ an effect reminiscent of a Higgs mechanism.  

To characterize the properties of this field-induced spin liquid, we also explore the effects of adding random non-magnetic impurities. We show that the spin liquid compensates the vacancy so that the magnetization remains independent of the dilution density for small densities $-$ reminiscent of a Meissner effect in a Higgs phase.


\section{Model and the ground state manifold}

\begin{figure}[h!]
    \centering
    \includegraphics[width=\columnwidth]{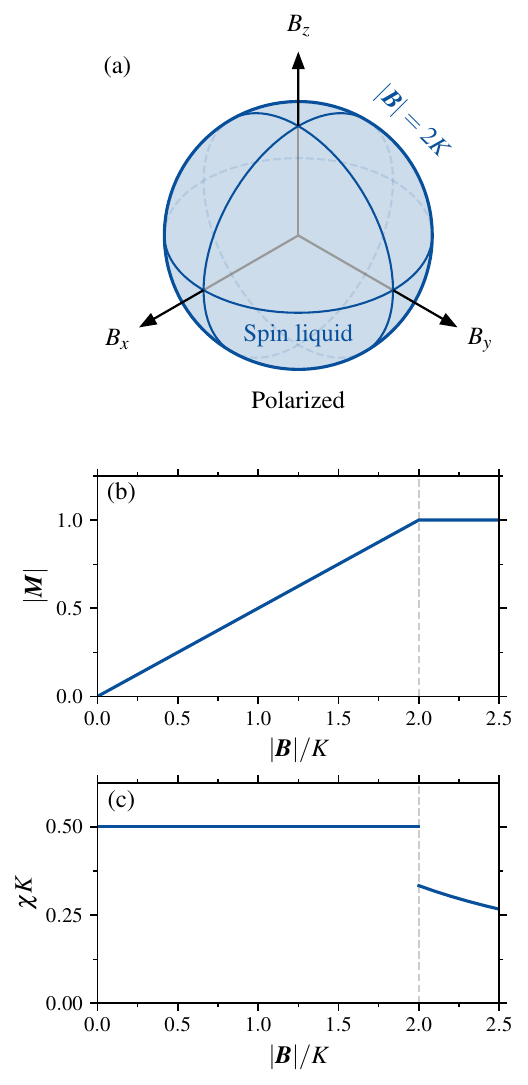}
    \caption{(a) Phase diagram of the classical anti-ferromagnetic Kitaev model in a magnetic field at zero temperature. For arbitrary field direction there is a field-induced spin liquid
    phase for field $|\vec{B}|\leq 2K$. (b) Magnetization as a function of field. The magnetization is linear in the spin liquid phase, with $\vec{M} = \vec{B}/(2K)$, until saturating at $|\vec{B}|=2K$.  (c) Average susceptibility $\chi \equiv \sum_{\mu}\chi_{\mu\mu}/3$ as a function of field is constant in the spin liquid phase and $\propto 1/|\vec{B}|$ in the polarized phase.}
    \label{fig:phase-diagram}
\end{figure}

The energy of the classical Kitaev honeycomb model in a magnetic field is given by
\begin{equation}
  E = K \sum_{\avg{ij} \in {\mu}} S^\mu_i S^\mu_j - \vec{B} \cdot \sum_i \vec{S}_i
  \label{eq:E_Kitaev}
\end{equation}
where $|\vec{S}_i|^2=1$ and $\mu=x,y,z$ labels the set of bonds with the particular exchange anisotropy.
We now look for the ground state energy $E_0$ as well as the set of states that
yield this minimum. We focus here on the anti-ferromagnetic case with $K>0$ in a magnetic field where 
the spin liquid phase survives to finite field. The ferromagnetic case is less rich, with the liquid phase being immediately destroyed, becoming
fully polarized at any finite magnetic field. We will briefly discuss this case at the end of this section.

To uncover the behavior of the anti-ferromagnetic case,  we first write the energy in Eq.~(\ref{eq:E_Kitaev}) as a sum over one sublattice of the honeycomb
\begin{equation}
  E = \sum_{i\in A}\sum_{\mu} \left[ K S^\mu_i S^\mu_{i+\mu} - B_\mu \left(S^\mu_i + S^\mu_{i+\mu}\right)\right],
    \label{eq:E_Kitaev_A}
\end{equation}
where $i+\mu \in B$ are the three neighbors of $i$ along the bond directions $\mu = x,y,z$. We note that $\sum_{i \in A} 1 = N/2$ where $N$ is the total number of sites. One can recast the expression above as a sum of squares 
\begin{align}
  E
&=  \frac{K}{2} \sum_{i\in A}\sum_{\mu}\left(S^\mu_i + S^\mu_{i+\mu} - \frac{B_\mu}{K}\right)^2
 -\frac{1}{2}\left(K+\frac{|\vec{B}|^2}{2K}\right)N.
\end{align}
Ground states thus satisfy the (linear) constraint
\begin{equation}
  S^\mu_i + S^\mu_{i+\mu} = \frac{B_\mu}{K}
  \label{eq:constraint}
\end{equation}
together with the (quadratic) constraint for the length,
\begin{equation}
  |\vec{S}_i|^2 = \sum_{\mu} (S^\mu_i)^2  = 1,
  \label{eq:constraint_lenght}
\end{equation}
and have minimal energy density 
\begin{equation}
    \frac{E_0}{N} =
- \frac{K}{2}- \frac{|\vec{B}|^2}{4K}.
\end{equation}
The local constraints and associated energy bound also imply a perfectly linear magnetization curve in the spin liquid phase, with
\begin{equation}
  \vec{M} \equiv  -\frac{1}{N}\frac{\del E_0}{\del \vec{B}} = \frac{\vec{B}}{2K}.
\end{equation}
This persists up to some critical field; since we must have $|\vec{M}|
\leq 1$, this implies a direction independent saturation field of $B_{s} =
2K$ after which one has $\vec{M} = \vhat{B}$. This can also be derived directly from the constraint equation, noting that $|S^\mu_i| \leq 1$. This implies that the constraints must be violated for $|\vec{B}| >2K$, with the energy density being $E_0/N = -K/2 - |\vec{B}|$. Similarly for $|\vec{B}| < 2K$ the susceptibility is isotropic and constant, with
\begin{equation}
\chi_{\mu\nu} \equiv \frac{\del M_\mu}{\del B_\nu} = \frac{\delta_{\mu\nu}}{2K}.
\label{eq:chi_liquid}
\end{equation}
At larger fields where $|\vec{B}|>B_s$ one finds the simple ``dipolar'' form
\begin{equation}
  \chi_{\mu\nu} = \frac{\delta_{\mu\nu}-\vhat{B}_\mu\vhat{B}_\nu}{|\vec{B}|}
\end{equation}
arising from the fully saturated moment. An identical susceptibility is found in the ferromagnetic case at any field strength. Note that there is a jump in the susceptibility at the saturation field, as shown in Fig.~\ref{fig:thermodynamics}. We show this magnetization curve and the susceptibility in Fig.~\ref{fig:phase-diagram}.

While not as straightforward to construct as in the zero-field case, when $|\vec{B}| < 2K$ one can find many states that satisfy this constraint [Eq.~(\ref{eq:constraint})] and thus saturate the bound on the energy, for example via classical Monte Carlo simulations supplemented with direct minimization techniques~\footnote{Note that the Cartesian states of \citet{baskaran2008} \emph{do not} satisfy this constraint; for generic field direction, the spin components associated with every bond must be non-zero.}.  The ground state phase diagram of this model is shown in Fig.~\ref{fig:phase-diagram}, with a classical spin liquid phase for $|\vec{B}|< 2K$ and a polarized phase
for $|\vec{B}|>2K$.

\subsection{Localized Zero Modes}
\label{sec:linearizedzeromodes}
\begin{figure}[b]
\begin{center}
\includegraphics[width=0.65\columnwidth]{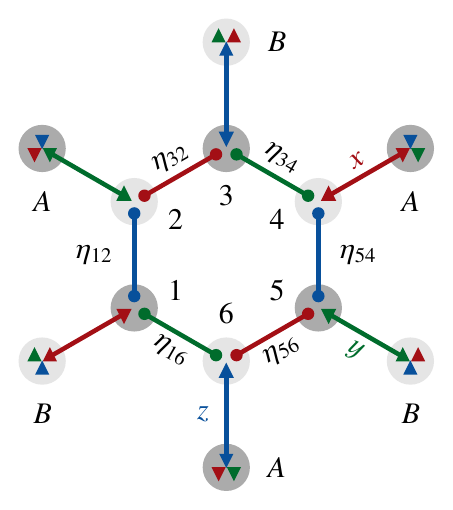}
\caption{A hexagon surrounded by a canted Néel state. Dark gray sites belong to the A sublattice, light gray to the B sublattice. The triangles inside the circles denote the corresponding (color-coded) components of the $\vec{S}_A$ and $\vec{S}_B$ Néel vectors. We allow for the small circles inside the big circles to take different values to find a weathervane mode. }
\label{fig:hexagon}
\label{axes} 
\end{center}
\end{figure}
For the simple ground states, the presence of large ground state degeneracy even at finite field can be seen directly. Perhaps the simplest ordered ground state is the two-sublattice Néel state, where spins on the A sublattice all point in the same direction $\vec{S}_A = (S_A^x,S_A^y,S_A^z)$ and on the B sublattice to $\vec{S}_B = (S_B^x,S_B^y,S_B^z)$. Inserting these spins into Eq.~(\ref{eq:E_Kitaev_A}), the energy of the N\'eel state is
\begin{equation}
  E_{{\rm N\acute{e}el}} = \frac{N}{2}\left[ K \, \vec{S}_A \cdot\vec{S}_B - \vec{B} \cdot (\vec{S}_A + \vec{S}_B) \right]
\end{equation}
for $N$ lattice sites. In the ground state, the spin configurations satisfy
\begin{equation}
  \vec{S}_A +  \vec{S}_B = \frac{\vec{B}}{K}
  \label{eq:SASBBconstraint}
\end{equation}
which can be achieved by canted configurations for $|\vec{B}| <2 K$. The extensive continuous ground state degeneracy can be diagnosed by the existence of zero energy local modes, such as (for example) the weathervane modes present in the kagome antiferromagnet~\cite{chalker1992,ritchey1993,shender1993}. 

To this end, we consider an isolated hexagon as illustrated in Fig.~\ref{fig:hexagon} with the remaining spins
fixed into the canted Néel state. Our goal will be to construct a localized deformation of the spins on this hexagon
that preserves the ground state constraints.
The surrounding canted Néel state pins, on each site of the hexagon, the spin component along the Kitaev axis of the bond connecting that site to the environment (indicated by the small triangles) via the constraints [Eq.~(\ref{eq:constraint})]: explicitly $S_1^x= - S^x_B + B_x/K = S_A^x$, and similarly $S_2^y=S_B^y$, following cyclic permutations around the loop.
The remaining components may vary, subject to the unit-length constraint $|\vec S_i|=1$.

We parametrize this constrained configuration space by bond variables $\eta_{ij}$ on the hexagon. Varying $\eta_{ij}$ preserves all pinned components and generates a deformation supported entirely on the hexagon:
\begin{subequations}
\label{eq:Si_eta}
\begin{align}
  \vec{S}_1 &= (S_A^x, S_1^y, S_1^z) = (S_A^x, S_A^y + \eta_{1,6}, S_A^z + \eta_{1,2})\,, \\
  \vec{S}_2 &= (S_2^x, S_B^y, S_2^z) = (S_B^x - \eta_{3,2}, S_B^y, S_B^z - \eta_{1,2})\,, \\
  \vec{S}_3 &= (S_3^x, S_3^y, S_A^z) = (S_A^x + \eta_{3,2}, S_A^y + \eta_{3,4}, S_A^z)\,, \\
  \vec{S}_4 &= (S_B^x, S_4^y, S_4^z) = (S_B^x, S_B^y - \eta_{3,4}, S_B^z - \eta_{5,4})\,, \\
  \vec{S}_5 &= (S_5^x, S_A^y, S_5^z) = (S_A^x + \eta_{5,6}, S_A^y, S_A^z + \eta_{5,4})\,, \\
  \vec{S}_6 &= (S_6^x, S_6^y, S_B^z) = (S_B^x - \eta_{5,6}, S_B^y - \eta_{1,6}, S_B^z)\,.
\end{align}
\end{subequations}
With this ansatz, Eq.~(\ref{eq:constraint}) is enforced on every bond of the hexagon, so the deformation satisfies the necessary ground-state condition.
The zero mode is realized as a one-parameter family $\{\eta_{ij}(\lambda)\}$ along which both the Kitaev exchange energy on the affected bonds and the Zeeman contribution remains invariant, yielding an exactly flat direction of the classical energy at finite field.

Imposing the unit-length constraints $|\vec{S}_i|^2=1$ for the spins above, and using $|\vec{S}_A|^2=|\vec{S}_B|^2=1$, we obtain a set of quadratic equations for the $\eta_{ij}$:
\begin{subequations}
\label{eq:hexagoneta}
\begin{align}
\eta_{1,6}^2 + 2 S_A^y \eta_{1,6} + \eta_{1,2}^2 + 2 S_A^z \eta_{1,2} &= 0\,, \label{eq:hexagoneta_a} \\
\eta_{3,2}^2 - 2 S_B^x \eta_{3,2} + \eta_{1,2}^2 - 2 S_B^z \eta_{1,2} &= 0\,, \label{eq:hexagoneta_b} \\
\eta_{3,2}^2 + 2 S_A^x \eta_{3,2} + \eta_{3,4}^2 + 2 S_A^y \eta_{3,4} &= 0\,, \label{eq:hexagoneta_c}\\
\eta_{3,4}^2 - 2 S_B^y \eta_{3,4} + \eta_{5,4}^2 - 2 S_B^z \eta_{5,4} &= 0\,, \label{eq:hexagoneta_d} \\
\eta_{5,6}^2 + 2 S_A^x \eta_{5,6} + \eta_{5,4}^2 + 2 S_A^z \eta_{5,4} &= 0\,, \label{eq:hexagoneta_e} \\
\eta_{5,6}^2 - 2 S_B^x \eta_{5,6} + \eta_{1,6}^2 - 2 S_B^y \eta_{1,6} &= 0\,. \label{eq:hexagoneta_f}
\end{align}
\end{subequations}

For small deviations, when the $\eta_{ij}$ are small, we can neglect $O(\eta^2)$ terms in Eqs.~(\ref{eq:hexagoneta}). What remains is a set of homogeneous linear equations with the following non-trivial solution
\begin{subequations}
\begin{align}
\eta_{1, 2} &= -\lambda S_B^x S_A^y & \eta_{3, 2} &= +\lambda S_A^y S_B^z & \eta_{3, 4} &= -\lambda S_A^x S_B^z \\
\eta_{5, 4} &= +\lambda S_A^x S_B^y & \eta_{5, 6} &= -\lambda S_B^y S_A^z & \eta_{1, 6} &= +\lambda S_B^x S_A^z
\end{align}    
\end{subequations}
where $\lambda \ll 1$ is a continuous parameter.
The existence of the nontrivial solution signals a zero mode at infinitesimal amplitude. 
However, for generic $\vec{S}_A, \vec{S}_B$, this mode does not extend to finite amplitude once quadratic terms are restored. We show in Appendix \ref{sec:local_mode}, including the quadratic terms in $\eta^2$, that it {\it does} extend to finite amplitude for a special class of backgrounds.

\subsection{Ferromagnetic Kitaev Exchange}
Like its antiferromagnetic counterpart ($K>0$), the classical ferromagnetic Kitaev model ($K<0$) is a spin liquid at zero field. But at any finite field the latter model is fully polarized. To see this, again we complete the square but now keeping the Zeeman term separate  
\begin{align}
  E
&=  \frac{K}{2} \sum_{i\in A}\sum_{\mu\in\{x,y,z\}} \left[ \left(S^\mu_i - S^\mu_{i+\mu} \right)^2 - B_\mu(S_i^\mu + S_{i+\mu}^\mu) \right]
 -\frac{K}{2}N.
\end{align}
The spin interaction term is minimized for $S_i^\mu = S_{i+\mu}^{\mu}$, which is identical to the manifold found in the antiferromagnetic case after application of a sublattice sign $\vec{S}_i \rightarrow (-1)^i \vec{S}_i$. This manifold therefore contains all fully polarized spin configurations. Since the fully polarized configurations along the field are the \emph{only} configurations that minimize the Zeeman term these are the configurations of the global minimum in the energy. So, whereas a classical spin liquid survives in some form at finite field for $K>0$, the ferromagnetic case $K<0$ is a simple polarized magnet for any $|\vec{B}|/K > 0$. 


\section{Thermodynamics}

\begin{figure*}[tp]
  \centering
  \includegraphics[width=\textwidth]{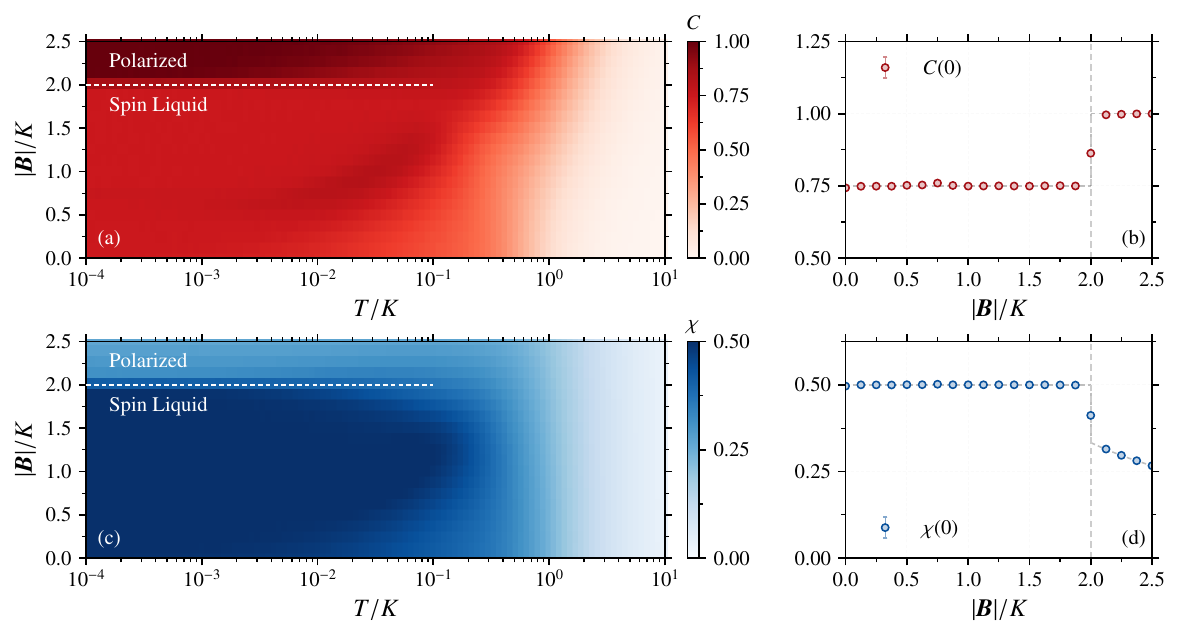}
  \caption{\label{fig:thermodynamics}
  Thermodynamics of the classical anti-ferromagnetic Kitaev model in an $[111]$ field.
  Monte Carlo simulations for a $L=24$ system with periodic boundary conditions taking approximately $10^6$ samples using heat-bath, over-relaxation and parallel tempering updates.
(a) Heat capacity shows a smooth function of temperature from the classical spin liquid regime into the polarized paramagnetic regime. (b) At low temperatures the heat capacity approaches $3/4$ in the spin liquid phase ($|\vec{B}|<2K$) before switching to $1$ in the polarized phase  ($|\vec{B}|>2K$). (c) The average susceptibility $\chi \equiv \sum_{\mu} \chi_{\mu\mu}/3$ is also featureless as a function of temperature, approaching a constant in the spin liquid phase at $T=0$. (d) At low temperature there is a discontinuity in the susceptibility at the critical field.
  }
\end{figure*}

Having established the nature of the ground states we now consider the finite temperature properties of the anti-ferromagnetic model at finite field. We focus here on the $[111]$ direction; other field directions are
similar (see Appendix~\ref{app:directions}). To study the finite temperature properties of this model we use
standard Monte Carlo techniques with local heat-bath and over-relaxation updates, combined with parallel tempering across temperatures.

Fig.~\ref{fig:thermodynamics} shows simulation results for the heat capacity and susceptibility as a function of temperature and field. Both quantities are smooth functions of the temperature, with no sign of a phase transition via order-by-disorder. Further, the behavior of the thermodynamic quantities is consistent with the nature of the liquid phase, as we will show more directly later.
The classical spin liquid regime thus apparently extends from zero field up to the threshold field. Density maps of the heat capacity and susceptibility further show that the only sharp feature is the threshold field  $|\vec{B}|=2K$. As expected, the magnetization increases linearly at low temperatures and the susceptibility converges to $\chi_{\mu\nu} = \delta_{\mu\nu}/(2K)$ as $T\rightarrow 0$ consistent with the result of Eq.~(\ref{eq:chi_liquid}) derived in the previous section from the constraint equations. 

The heat capacity per spin tends to $3/4$ at zero temperature in the spin liquid regime and $1$ in the polarized paramagnetic phase. The latter is the standard expectation from classical equipartition with the full $2$ degrees of freedom per site each contributing a heat capacity $1/2$. The departure of this value in the spin liquid regime is a consequence of a manifold of ground states with a large number of zero modes. Counting degrees of freedom and constraints~\cite{moessner1998a}, there are $3N$ spin variables with $N$ length constraints and $N/2$ ground state constraints. There are thus $N/2$ degrees of freedom in the ground state manifold and $3N/2$ modes that are not. By equipartition this yields a heat capacity of 
\begin{equation}
C(T\rightarrow 0) = \frac{3N}{2} \times \frac{1}{2} + \frac{N}{2} \times 0 = \frac{3N}{4}
\label{eq:CT0}
\end{equation}
total, or $3/4$ per spin as required (the density of quartic or ``soft'' modes vanishes). This constraint counting is consistent with analysis of the normal modes about generic ground states, as was found (e.g.) at zero field by \citet{baskaran2008} where spin waves out of some special ground states (self-avoided walk states) were analyzed.

\section{Correlations}
With the thermodynamic properties understood, we look to the correlations in the
spin liquid phase. We will consider both spin and quadrupolar correlations. 

At zero-field the plaquette symmetries enforce short range spin correlations~\cite{baskaran2008}. 
The spin correlations remain short-ranged at finite field, but their character changes as the
field strength increases. This can be inferred from the constraint, with
\begin{equation}
\label{eq:spin-correlation}
  \avg{S^\mu_iS^\mu_{i+\mu}} = \frac{B_\mu^2}{2K^2} - \avg{(S^\mu_i)^2}
\end{equation}
The average nearest neighbour correlator is simply
\begin{equation}
  \sum_\mu \avg{S^\mu_iS^\mu_{i+\mu}} = \frac{|\vec{B}|^2}{2K^2} - 1.
\end{equation}
Note that this changes sign from negative to positive when $|\vec{B}| = \sqrt{2} K$, independent of field direction. This is broadly consistent with going from the antiferromagnetic Kitaev limit at zero field to the field polarized ferromagnet at $|\vec{B}|=2K$.

\begin{figure*}[tp]
  \centering
  \includegraphics[width=\textwidth]{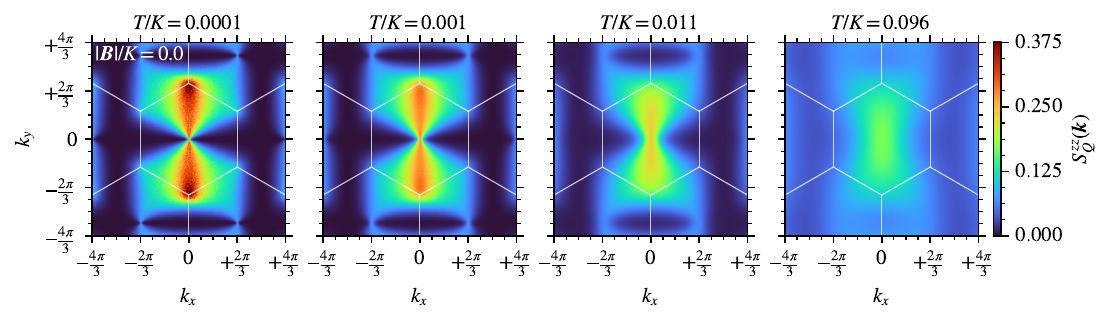}
  \caption{\label{fig:pinch-point-zero-field}
    Appearance of a pinch point at $\vec{k}=0$ in the spin-quadrupole
correlator, $S^{zz}_{Q}(\vec{k})$ for the anti-ferromagnetic Kitaev model at zero field as temperature $T$ is lowered. Monte Carlo simulations for a $L=120$ system with periodic boundary conditions over $10^6$ sweeps using heat-bath, over-relaxation and parallel tempering updates, taking samples of the quadrupolar structure factor every $10$ sweeps.
  }
\end{figure*}

\begin{figure}[tp]
  \centering
  \includegraphics[width=\columnwidth]{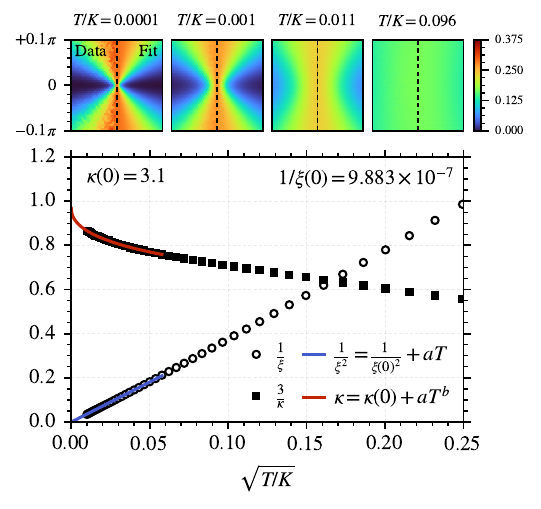}
  \caption{\label{fig:pinch-point-zero-field-scaling}
    Scaling of the pinch point width in the spin-quadrupole
    structure factor in the Kitaev model at zero field.
    System size is $L=120$ and was run for $10^6$ sweeps.
    For each temperature the wave-vector dependence of $S^{zz}_Q(\vec{k})$ 
    is fit over a range of wave-vectors
    near $\vec{k}=0$ using Eq.~(\ref{eq:finite-temp-pp}) (see upper plots). The temperature dependence of $1/\xi(T)$ is then
    fit to $1/\xi^2 =  {1/\xi(0)^2 + aT}$ yielding
    $1/\xi(0) \approx 10^{-6}$. The temperature dependent stiffness $\kappa$ is
    fit to $\kappa = \kappa(0) + a T^b$, yielding
    $\kappa(0) \approx 3.1$ 
  }
\end{figure}

The quadrupolar degrees of freedom are more interesting and show algebraic correlations at zero-field~\cite{chandra2010}. 
A natural definition of on-site spin quadrupoles is in terms of a traceless tensor
\begin{equation}
\mat{Q}_i \equiv \left(\begin{array}{ccc}
     (S^x_i)^2 -\frac{1}{3} & S^x_i S^z_i & S^x_i S^z_i  \\
     S^y_i S^z_i & (S^y_i)^2 -\frac{1}{3} &  S^y_i S^z_i  \\
     S^z_i S^x_i & S^z_i S^y_i & (S^z_i)^2 -\frac{1}{3} 
\end{array}
\right)
\end{equation}
where ${\rm Tr}({\mat{Q}_i}) = |\vec{S}_i|^2-1=0$ due to the length constraint.
For our purposes we only need the diagonal components of this tensor and so we will define
the three quadrupolar moments~\cite{chandra2010}
\begin{equation}
Q^{\mu}_i \equiv Q^{\mu\mu}_i \equiv (S^\mu_i)^2 -\frac{1}{3}.
\end{equation}
We  study these correlations in momentum space, defining the Fourier transform
\begin{equation}
Q^{\mu}_{\vec{k}} \equiv \frac{1}{\sqrt{N}} \sum_i e^{-i\vec{k}\cdot\vec{r}_i} Q^{\mu}_i
  \label{eq:Qk}
\end{equation}
where $\vec{r}_i$ is the position of lattice site $i$. The quadrupolar structure factor $S_{Q}(\vec{k})$
can then be defined as
\begin{equation}
S^{\mu\nu}_{Q}(\vec{k}) \equiv 
\frac{1}{N}\sum_{ij} e^{i\vec{k}\cdot(\vec{r}_i-\vec{r}_j)} \avg{Q^{\mu}_i Q^{\nu}_j} = 
\braket{{Q}^{\mu}_{-\vec{k}} Q^{\nu}_{\vec{k}}}.
\end{equation}
We show the $S^{zz}_{Q}(\vec{k})$ quadrupolar structure factor from our Monte Carlo simulations at zero field
in Fig.~\ref{fig:pinch-point-zero-field}. We see that a pinch point develops as $T \rightarrow 0$
with the pinch point width scaling as $\propto \sqrt{T}$, as shown in Fig.~\ref{fig:pinch-point-zero-field-scaling}.

\begin{figure*}[tp]
  \centering
  \includegraphics[width=\textwidth]{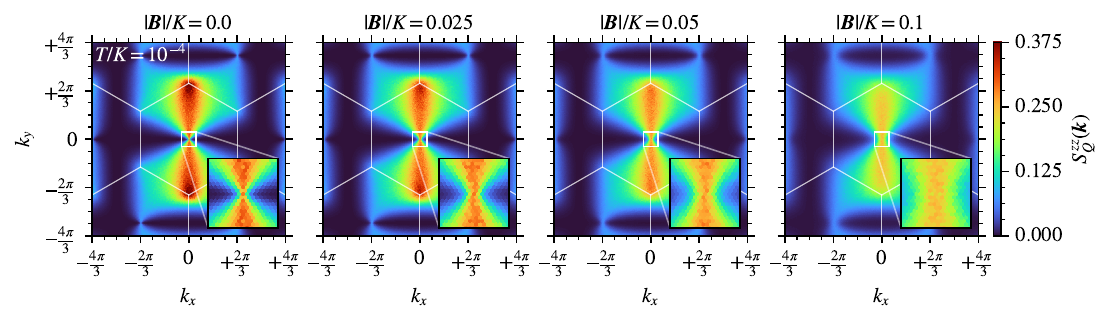}
  \caption{\label{fig:pinch-point-finite-field}
    Spin-quadrupole correlator, $S^{zz}_{Q}(\vec{k})$, for the anti-ferromagnetic Kitaev model at fixed temperature $T/K=10^{-4}$ with field along $[111]$ and field strengths $|\vec{B}|/K = 0, 0.025,0.05,0.1$. The pinch point present at $\vec{k}=0$ broadens as the field increases (see insets).  Monte Carlo simulations for a $L=120$ system with periodic boundary conditions over $10^6$ sweeps using heat-bath, over-relaxation and parallel tempering updates, taking samples of the quadrupolar structure factor every $10$ sweeps.
  }
\end{figure*}

\begin{figure}[tp]
  \centering
  \includegraphics[width=\columnwidth]{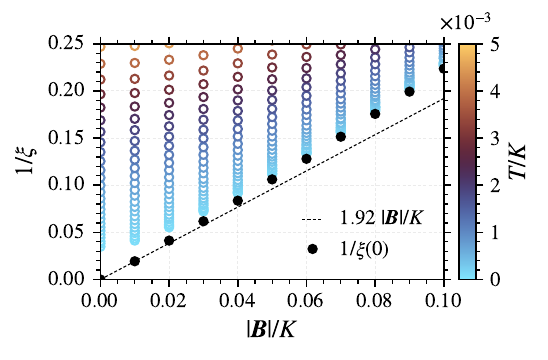}
  \caption{\label{fig:pinch-point-width-scaling}
    Scaling of the pinch point width as function of $[111]$ field
    at low temperature. For each field value we extrapolate
    to $T=0$ by fitting $1/\xi^2 = 1/\xi(0)^2+aT$ to several temperatures in the range $10^{-4} \leq T/K \leq 10^{-3}$. The width follows
    a linear scaling with field and is compared to the result from the coarse-grained theory, $\sim 1.92|\vec{B}|/K$.
     Monte Carlo simulations for a $L=120$ system with periodic boundary conditions over $10^6$ sweeps using heat-bath, over-relaxation and parallel tempering updates, taking samples of the quadrupolar structure factor every $10$ sweeps.
  }
\end{figure}

We can examine the same pinch points in the finite field case. We show the quadrupolar structure factor
for a $[111]$ field at $T/K=10^{-4}$ in Fig.~\ref{fig:pinch-point-finite-field}. We see that as $|\vec{B}|$ is increased the pinch-point broadens and 
acquires a finite width. Quantitative statements can be made by
extracting the pinch point width (extrapolated to $T=0$) as a function of field, as is shown in Fig.~\ref{fig:pinch-point-width-scaling}. We see that the width scales $\propto |\vec{B}|/K$, growing linearly with field at low fields. The absence of the pinch points in the finite field spin liquid is a generic feature of this phase, independent of field direction and strength $|\vec{B}|<2K$.

\section{Coarse-grained Theory}
We now show that these results can be understood using an extension
of the coarse-grained picture of \citet{chandra2010}. First, we review the coarse-grained picture for
the quadrupolar liquid in the zero field limit as a Coulomb phase. We then show that the effect of a field can be interpreted as the introduction of fluctuating charges in this Coulomb phase.
\subsection{Zero field}
First note that the normalization of the spins imposes the 
constraint on the quadrupoles $\sum_{\mu} Q^{\mu}_i =|\vec{S}_i|^2-1 =0$.
One can cast the spin quadrupoles in terms of two fields ${E}_{ij}$ and $F_{ij}$
defined on the links of the honeycomb lattice
\begin{align}
    F_{ij} &=  \frac{1}{2}\left(Q^{\mu}_{j} - Q^{\mu}_i\right), \\
    E_{ij} &= \frac{1}{2} \epsilon_i \left(Q^{\mu}_{j} + Q^{\mu}_i\right),
\end{align}
where the bond $\avg{ij}$ is of type $\mu=x,y,z$ and
the sign factor $\epsilon_i =+1$ on the $A$ sublattice and $\epsilon_i =-1$ on the $B$
sublattice. As defined these are antisymmetric 
$E_{ij} = -E_{ji}$ and $F_{ij}=-F_{ji}$ as expected for link fields.

If the spin configuration is a ground state of the antiferromagnetic Kitaev 
model with have that $S^{\mu}_i =-S^{\mu}_{i+\mu}$ and thus
\begin{equation}
Q^{\mu}_i = Q^{\mu}_{i+\mu}
\end{equation}
This condition forces $F_{ij}=0$ and we thus consider only the $E_{ij}$.
The divergence $E_{ij}$ is directly related to this constraint. If we define
\begin{align}
{\rm div}_i(E) &\equiv \sum_{\mu} E_{i,i+\mu},    
\end{align}
then in terms of the spin quadrupoles 
\begin{align}
{\rm div}_i(E) &= \frac{1}{2} \epsilon_i\sum_{\mu} \left(Q^{\mu}_{i+\mu}+Q^{\mu}_{i}\right)= \frac{1}{2} \epsilon_i\sum_{\mu} Q^{\mu}_{i+\mu},
\end{align}
where we have fixed the first sum in each using the spin normalization.
For Kitaev ground states we thus have
\begin{align}
{\rm div}_i(E) &=0.
\end{align}
We thus see that in the ground state manifold the spin quadrupoles can
be recast as a divergence free vector field on the honeycomb lattice.

To understand the behavior of this divergence-free field we define
a coarse-grained version $\vec{E}(\vec{r})$ that we assume is slowly
changing relative to lattice scales. Explicitly,
\begin{align}
E_{i,i+\mu} &\equiv \bar{a} \epsilon_i  \vhat{d}_{\mu} \cdot \vec{E}(\vec{r}_i + a\epsilon_i \vhat{d}_{\mu}/2),    
\end{align}
where the $\vec{d}_{\mu}$ are the directions of the nearest-neighbour
bonds on the $A$ sublattice. The nearest neighbour distance is denoted $a$ and
$\bar{a} = \sqrt{3}a$ is the side-length of the Voronoi cell of each site -- this choice
of normalization simplifies the statement of Gauss' law in the presence of charges.
The sublattice sign structure ensures $E_{i,i+\mu}=-E_{i+\mu,i}$.

In the long wave-length limit the lattice divergence is proportional to
the continuum divergence
\begin{align}
\sum_{\mu}E_{i,i+\mu} &= 
\bar{a}\epsilon_i
\sum_{\mu}
\vhat{d}_{\mu} \cdot \vec{E}(\vec{r}_i + a\epsilon_i \vhat{d}_{\mu}/2)
\approx
\frac{3a\bar{a}}{4}(\grad \cdot \vec{E})(\vec{r}_i),
\end{align}
where we have used that $\sum_{\mu} \vhat{d}_{\mu}=0$ and $\sum_{\mu}
\hat{d}^a_{\mu} {d}^b_{\mu} = 3\delta_{ab}/2$. Note that the area occupied
by a single site is $A_1 = A/N = 3a\bar{a}/4 = 3\sqrt{3}a^2/4$. We can then write
\begin{align}
{\rm div}_i(E) = \sum_{\mu}E_{i,i+\mu} &\approx A_1 (\grad \cdot \vec{E})_i    
\end{align}

The simplest coarse grained theory, valid at $T=0$, would involve only $\vec{E}(\vec{r})$. 
As in other Coulomb phases~\cite{Huse2003,henley2010coulomb},  we consider the entropic contribution to the free energy
\begin{equation}
    \beta \mathcal{F} = \frac{\mathcal{K}}{2} \int d^2 r |\vec{E}(\vec{r})|^2,
    \label{eq:coarse-grained}
\end{equation}
subject to the constraint $\grad \cdot \vec{E}=0$. The parameter $\mathcal{K}$ is a
phenomenological constant characteristic of this ground state manifold.
In Fourier space  we can define
\begin{align}
\vec{E}(\vec{k}) &\equiv \frac{1}{\sqrt{A}} \int d^2r e^{-i\vec{k}\cdot\vec{r}} \vec{E}(\vec{r}), &
    \vec{E}(\vec{r}) &\equiv \frac{1}{\sqrt{A}} \sum_{\vec{k}} e^{i\vec{k}\cdot\vec{r}} \vec{E}(\vec{k}).
\end{align}
This yields the effective free energy
\begin{equation}
\beta \mathcal{F} = \frac{\mathcal{K}}{2} \sum_{\vec{k}} |\vec{E}(\vec{k})|^2,
\label{eq:coarse-grained-fourier}
\end{equation}
with the divergence free constraint recast as $\vec{k}\cdot\vec{E}(\vec{k}) = 0$. 
Writing this field in terms of an orthogonal basis $\vhat{k}$ and $\vhat{p} \equiv \hat{k}_y \vhat{x} - \hat{k}_x \vhat{y}$ at each wave-vector
\begin{equation}
\vec{E}(\vec{k}) = \vhat{k}E_{||}(\vec{k}) + \vhat{p} E_{\perp}(\vec{k})
\end{equation}
the constraint requires that the longituidinal component vanishes, $E_{||}(\vec{k})=0$. The free energy then only
involves the transverse components
\begin{equation}
\beta \mathcal{F} = \frac{\mathcal{K}}{2} \sum_{\vec{k}} |E_{\perp}(\vec{k})|^2.
\end{equation}
This is a free Gaussian theory and therefore the correlations are
$
\avg{E_{\perp}(-\vec{k})E_{\perp}(\vec{k})} = {1}/{\mathcal{K}}.
$
Using   $\vec{E}(\vec{k}) = \vhat{p} E_{\perp}(\vec{k})$ the
correlations of the full field are
\begin{equation}
\avg{E_{\mu}(-\vec{k})E_{\nu}(\vec{k})} = 
\frac{\hat{p}_{\mu}\hat{p}_{\nu}}{\mathcal{K}} = \frac{1}{\mathcal{K}}\left(\delta_{\mu\nu} - \hat{k}_{\mu}\hat{k}_{\nu}\right).
\end{equation}
This coarse grained field thus has pinch-point correlations near $\vec{k}=0$.

These pinch point correlations manifest directly in the quadrupolar structure factor.
Using the constraint $Q^{\mu}_i = \epsilon_i E_{i,i+\mu}$ and thus
\begin{equation}
Q^{\mu}_{\vec{k}}= \frac{1}{\sqrt{N}}\sum_i e^{-i\vec{k}\cdot\vec{r}_i} \epsilon_i E_{i,i+\mu}.
\end{equation}
In terms of the coarse-grained fields this is given by
\begin{equation}
Q^{\mu}_{\vec{k}}= \frac{\bar{a}}{\sqrt{N}}\sum_i e^{-i\vec{k}\cdot\vec{r}_i} \vhat{d}_{\mu} \cdot \vec{E}(\vec{r}_i + a\epsilon_i \vhat{d}_{\mu}/2).
\end{equation}
Assuming the field $\vec{E}(\vec{r})$ varies slowly in space the shift
of the argument can be neglected, $\vec{E}(\vec{r}_i + \epsilon_i \vhat{d}_{\mu}/2) \approx \vec{E}(\vec{r}_i)$, and thus
\begin{equation}
Q^{\mu}_{\vec{k}} \approx \bar{a}\sqrt{\frac{N}{A}} \vhat{d}_{\mu} \cdot \vec{E}(\vec{k}).
\end{equation}
The quadrupolar degrees of freedom are thus directly proportional to the coarse-grained electric field.
The quadrupolar structure factor is then
\begin{equation}
S^{\mu\nu}_Q(\vec{k}) \approx  \frac{\bar{a}^2 N}{A} \trp{\vhat{d}_{\mu}} \avg{{\vec{E}(-\vec{k})} \trp{\vec{E}(\vec{k})}} \vhat{d}_{\nu}
\end{equation}
The coarse-grained theory then implies that
\begin{equation}
S^{\mu\nu}_Q(\vec{k}) \approx \frac{1}{\kappa} \left[\vhat{d}_{\mu}\cdot \vhat{d}_{\nu} - (\vhat{k}\cdot\vhat{d}_{\mu})(\vhat{k}\cdot\vhat{d}_{\nu})\right]
\end{equation}
where $\kappa \equiv \mathcal{K}A/(N\bar{a}^2) = \sqrt{3} \mathcal{K}/4$.
The diagonal components exhibit pinch points with
\begin{equation}
S^{\mu\mu}_Q(\vec{k})\approx \frac{1}{\kappa} \left[1 - (\vhat{k}\cdot\vhat{d}_{\mu})^2\right]
\end{equation}
For $\vhat{k}$ parallel to $\vhat{d}_{\mu}$ this is zero, while for $\vhat{k}$ perpendicular to $\vhat{d}_{\mu}$ this is equal to $1/\kappa$. At finite
temperature we expect these pinch points to acquire a temperature dependent
width $1/\xi(T)$ and stiffness $\kappa(T)$. 
Explicitly,
\begin{equation}
\label{eq:finite-temp-pp}
S^{\mu\mu}_Q(\vec{k})\approx  \frac{1}{\kappa(T)}\left(
1-\frac{
(\vec{k}\cdot \vhat{d}_{\mu})^2
}{|\vec{k}|^2+1/\xi(T)^2}
\right)
\end{equation}
where as $T\rightarrow 0$ we have $1/\xi(T) \rightarrow 0$ and
$\kappa(T) \rightarrow \kappa$.

This is precisely the pinch point structure observed in $S^{\mu\mu}_{Q}(\vec{k})$ in our Monte Carlo simulations at zero-field. 
The values of the parameters $\kappa$ and $\xi$ can be extracted from these simulations by fitting the numerical data to
the expected result from the coarse-grained theory [Eq.~(\ref{eq:finite-temp-pp})] over a small window of wave-vectors near $\vec{k}=0$.
We show this fitting in Fig.~\ref{fig:pinch-point-zero-field-scaling}, using a fit over a low temperature range to extrapolate to $T=0$.
As expected for a pinch-point the width of the Lorentzian vanishes as $T\rightarrow 0$, becoming infinitely sharp. For the stiffness, extrapolation to $T=0$ yields $\kappa \approx 3.1$, though the precise value depends somewhat on the fitting function used.

\subsection{Layer argument}
\begin{figure}[tp]
  \centering
  \includegraphics[width=0.95\columnwidth]{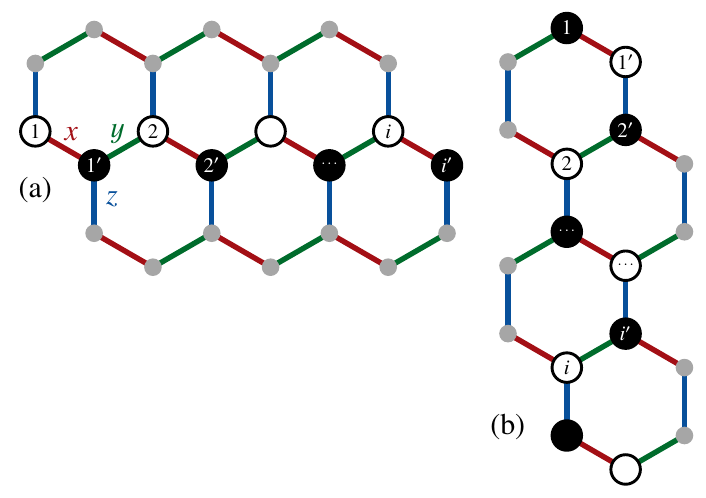}
  \caption{
  \label{fig:layer_argument}
Figure to illustrate the layer argument for the pinch point in momentum space in zero field. Adjacent layers on the honeycomb lattice are labelled by unprimed and primed indices. The vertical bonds are $z$ bonds. 
(a) For layers perpendicular to the $z$ bonds the sum $\sum_{i\in\mathcal{L}} Q_i^z = {\rm const.} $ leading to a pinch point in the corresponding quadrupolar correlation function. 
(b) In layers perpendicular to the $x$ bond, the  $\sum_{i\in\mathcal{L}} Q_i^x = {\rm const.} $ holds and the associated pinch point is rotated by 120$^\circ$ with respect to the case in (a).
  }
\end{figure}

The origin of these pinch point correlations can also be understood by a more direct real-space argument. Consider a pair of adjacent ``layers'' $\mathcal{L}$ and $\mathcal{L}'$ of spins from sublattice $A$ as illustrated in Fig.~\ref{fig:layer_argument} by primed and unprimed labels. If we sum the squares of the component of the spin with associated bonds
perpendicular to the layer (say $z$) we have
\begin{equation}
\sum_{i \in \mathcal{L}} (S_i^z)^2 = 
\sum_{i \in \mathcal{L}} \left(1-(S_i^x)^2-(S_i^y)^2\right)
\end{equation}
due to length normalization. Now the ground state constraint requires that for
each $i \in \mathcal{L}$ there is a neighbouring $i+x \in \mathcal{L}'$ such that
$(S^x_i)^2 = (S^x_{i+x})^2$ and a neighbouring $i+y \in \mathcal{L}'$ such that
$(S^y_i)^2 = (S^y_{i+y})^2$. We thus have
\begin{align}
\sum_{i \in \mathcal{L}} \left( S_i^z \right)^2 &= 
\sum_{i \in \mathcal{L}} \left(1-(S_{i+x}^x)^2-(S_{i+y}^y)^2\right) \nonumber \\
&= \sum_{i \in \mathcal{L}'} \left(1-(S_{i}^x)^2-(S_{i}^y)^2\right) = \sum_{i \in \mathcal{L}'} (S^z_i)^2
\end{align}
This sum $\sum_{i \in \mathcal{L}}(S^z_i)^2$ is thus constant and independent of the
layer. This perfect correlation implies that the Fourier transform of the correlation
function of $Q^z_i$ is non-zero when passing through $\vec{k}=0$ along the direction
perpendicular to the layers. However, while the sum is constant, $(S_i^z)^2$ itself is disordered within each layer so the correlator vanishes when approaching from the perpendicular direction. This defines a pinch point in the quadrupolar correlation
at $\vec{k}=0$. A similar argument can be made for other spin components (see Fig.~\ref{fig:layer_argument}).

\subsection{Non-zero field}
We now extend this coarse-grained theory to finite magnetic field. 
Focusing on the $E_{ij}$ field, when $\vec{B} \neq 0$ the divergence
is non-zero
\begin{equation}
{\rm div}_i(E) = \frac{1}{2} \epsilon_i\sum_{\mu} Q^{\mu}_{i+\mu} \equiv q_i,
\end{equation}
and we interpret the $q_i$ as fluctuating charges.
These charges $q_i$ can be expressed in terms of the spins using the constraint 
[Eq.~(\ref{eq:constraint})]
\begin{equation}
q_i = \frac{1}{2}\epsilon_i \left( \frac{|\vec{B}|^2}{K^2} - \frac{2\vec{B}\cdot \vec{S}_i}{K}\right).
\end{equation}
Since $\avg{\vec{S}_i} = \vec{B}/(2K)$ we have  zero net charge on average $\avg{q_i}=0$.
However, there are still fluctuations and correlations between these charges.
We can write
\begin{equation}
\avg{q_i q_j} =  \epsilon_i \epsilon_j \left( 
\sum_{\mu\nu} \frac{B_{\mu} B_{\nu}}{K^2} \avg{S^{\mu}_i S^{\nu}_j}
-\frac{
|\vec{B}|^4}{4K^4} 
\right)
\end{equation}
Since at zero-field the spin correlations are ultra-short ranged, we expect that 
at small field $|\vec{B}| \ll K$, the correlations between charges will be confined
to on-site or neighboring sites. Explicitly [see also Eq.~(\ref{eq:spin-correlation})]
\begin{subequations}
\begin{align}
    \avg{(S^{\mu}_i)^2} &= +\frac{1}{3} + O(|\vec{B}|^2), \\
    \avg{S^{\mu}_i S^{\mu}_{i+\mu}} &= \frac{1}{2}\left[ \frac{B_{\mu}^2}{K^2}-\avg{(S^{\mu}_{i+\mu})^2}-\avg{(S^{\mu}_i)^2}\right] \nonumber \\
&= -\frac{1}{3} + O(|\vec{B}|^2),
\end{align}
\end{subequations}
where we have kept only the leading term in powers of the field~\footnote{
Note that for a $[111]$ field $\protect\langle (S^{\mu}_i)^2 \protect\rangle = 1/3$ is exact as it is
required by the combination of normalization and three-fold symmetry.
}.
The fluctuating charges thus have the correlations
\begin{equation}
\avg{q_i q_j} = \frac{|\vec{B}|^2}{3K^2} \delta_{ij} +  
\sum_{\mu}\frac{B_{\mu}^2}{3K^2} \delta_{j,i+\mu} \equiv C_{ij},
\label{eq:charge-correlations}
\end{equation}
where we have used that $\epsilon_i \epsilon_{i+\mu}=-1$.

We now can add fluctuating charges to our coarse grained theory
with this modified constraint.  Augment the zero-field
free energy [Eq.~(\ref{eq:coarse-grained})] a term
that reproduces the above charge correlations
\begin{equation}
\beta\mathcal{F} = \frac{\mathcal{K}}{2}\int d^2 r |\vec{E}(\vec{r})|^2 +\frac{1}{2} \sum_{ij} C^{-1}_{ij} q_i q_j +\cdots,
\end{equation}
where $C^{-1}_{ij}$ are the elements of the inverse of the matrix $C_{ij}$.
To coarse-grain the constraint, we write the charge $q_i$ in 
terms of a charge density $\rho(\vec{r}_i) = q_i/A_1$.
With this definition the constraint on the divergence of $\vec{E}$
is a familiar Gauss' law $\grad \cdot \vec{E} = \rho$.
For the free energy it will be simplest work in Fourier space where
\begin{equation}
\beta\mathcal{F} = \frac{\mathcal{K}}{2} \sum_{\vec{k}} |\vec{E}(\vec{k})|^2+\frac{1}{2} \sum_{\vec{k}} C^{-1}_{\vec{k}} |q_{\vec{k}}|^2 +\dots
\end{equation}
The coarse-grained charge density is related to $q_{\vec{k}}$ via
\begin{equation}
\rho(\vec{k}) = \int \frac{d^2 r}{\sqrt{A}} e^{-i\vec{k}\cdot\vec{r}}\rho(\vec{r}) 
\approx \frac{q_{\vec{k}}}{\sqrt{A_1}}.
\end{equation}
The complete coarse-grained theory, including the electric field and the fluctuating
charges is then
\begin{equation}
\beta \mathcal{F} = \frac{\mathcal{K}}{2} \sum_{\vec{k}} |\vec{E}(\vec{k})|^2
+\frac{1}{2} \sum_{\vec{k}} \mathcal{C}(\vec{k})^{-1}|\rho(\vec{k})|^2 +\dots,
\end{equation}
where $ \mathcal{C}(\vec{k})^{-1} \equiv A_1 C^{-1}_{\vec{k}}$ which is subject
to the constraint $\vec{k} \cdot \vec{E}(\vec{k}) = \rho(\vec{k})$. Integrating out the $\rho(\vec{k})$ field is trivial using the constraint, yielding
\begin{equation}
\beta \mathcal{F} = \frac{\mathcal{K}}{2} \sum_{\vec{k}} |\vec{E}(\vec{k})|^2
+\frac{1}{2} \sum_{\vec{k}} \mathcal{C}(\vec{k})^{-1}|\vec{k}\cdot\vec{E}(\vec{k})|^2 +\dots.
\end{equation}
In the long-wavelength limit  only $\mathcal{C}(\vec{0}) \equiv \sigma^2$ is relevant. Using the explicit
form for the matrix $C_{ij}$ [Eq.~(\ref{eq:charge-correlations})] one can show that
\begin{equation}
\sigma^2 = \frac{2|\vec{B}|^2}{3A_1 K^2},
\end{equation}
which includes contributions from both the on-site and nearest-neighbor correlations of the charges.
The correlations of $\vec{E}(\vec{k})$ are then
\begin{equation}
\avg{E_{\mu}(-\vec{k})E_{\nu}(\vec{k})} =  \frac{1}{\mathcal{K}}\left(
\frac{(|\vec{k}|^2+\xi^{-2})\delta_{\mu\nu} - 
|\vec{k}|^2\hat{k}_{\mu}\hat{k}_{\nu}
}{|\vec{k}|^2+\xi^{-2}},
\right)
\end{equation}
where the correlation length $\xi$ is defined as
\begin{equation}
\frac{1}{\xi} = \sigma \sqrt{\mathcal{K}} = 
 \sqrt{\frac{32\kappa}{27}} \frac{|\vec{B}|}{a K}.
\end{equation}
We thus see that the fluctuating charges induced by the finite magnetic
field broaden the pinch-points, and the pinch-point width is expected to
scale linearly in the field strength. 

Using value for $\kappa \approx 3.1$ extracted from the zero-field structure factor
we can estimate the coefficient of this width as
\begin{equation}
\frac{1}{\xi} \approx 1.92 \frac{|\vec{B}|}{aK}
\end{equation}
This result matches well with the scaling of the pinch point width (extrapolated to $T=0$) shown in Fig.~\ref{fig:pinch-point-width-scaling}. Note that this derivation has a number of elements that are somewhat ``ad-hoc'' and would likely prevent it from
being truly quantitative, such as the neglect of the $F_{ij}$ variables at finite field, and the neglect of any field-dependent
variation in the stiffness $\kappa$. However, we believe the physical mechanism presented here correctly describes the destruction
of the quadrupolar Coulomb phase upon introduction of a finite field.


\section{Dilution and perfect screening}

To explore this field induced spin liquid state further, we now consider the effect of dilution, removing spins \cite{villain1979insulating,henley2001}. In other classical spin liquids the response of
the liquid to the diluted spin can reveal characteristics of the correlated state, for example by binding a fractional magnetic moment to the diluted site~\cite{Moessner1999,Sen2011,sen2012,Rehn2016,flores2024}. 

In the case of the classical Kitaev model, we find that the constraints can be preserved
and the spin liquid remains intact up to a reduced threshold field even in the presence of diluted sites. Though one might expect that the magnetization would be reduced once a single moment is removed,
we find -- counter-intuitively -- that the total magnetization remains unchanged. The remaining moments {\it compensate exactly} for the absence of an isolated spin, with the
average magnetization increasing near the diluted site and balancing the lost site -- in other words, this is {\it ``perfect compensation''}. 

With the behavior of a single spin understood we then proceed to remove more spins. We find that while the threshold field remains finite for two missing spins the simplest configurations where the threshold field is zero involve three vacancies that decouple one spin from the rest. With these results in hand, we discuss the nature of the spin liquid obtained by randomly diluting the system with vacancy probability $\rho$.

To consider diluted sites, we revisit the arguments used to derive the constraints with the substitution $\vec{S}_i
\rightarrow n_i \vec{S}_i$ where $n_i = 0$ or $1$.  One
finds
\begin{multline}  
  E 
=  \frac{K}{2} \sum_{i\in A}\sum_{\mu}\left(n_i S^\mu_i + n_{i+\mu}S^\mu_{i+\mu} - \frac{B_\mu}{K}\zeta^\mu_i\right)^2  \\ 
-\frac{N}{2}\left[(1-\rho)K + (1-\rho_2)\frac{|\vec{B}|^2}{2K}\right].
\end{multline}
We have defined $\zeta^\mu_i \equiv n_i+n_{i+\mu}-n_in_{i+\mu}$ which
is zero if $n_i=n_{i+\mu}=0$ and one otherwise -- in other words it is zero only for
bonds where both $i$ and $i+\mu$ have been removed. 
The total number of these bonds of type $\mu$ is $\sum_i \zeta^\mu_i = (1-\rho_2^\mu)N$ where $\rho_2^\mu$ is the density of missing $\mu$ bonds. 
For simplicity, we assume
that $\rho^\mu_2 \equiv \rho_2$ does not depend on the bond type.
The number of non-diluted spins is given by $N_s \equiv \sum_i n_i \equiv
(1-\rho) N$ where $0 \leq \rho \leq 1$ is the density of diluted sites
(with respect to the original lattice).  

One can thus see that any
states that satisfy
\begin{equation}
 n_i S^\mu_i +n_{i+\mu} S^\mu_{i+\mu} = \frac{B_\mu}{K}\zeta^{\mu}_i,
\end{equation}
are minimal energy states of the diluted interacting system.
For a pair of sites with
$n_i = n_{i+\mu} = 1$, these constraints are the same as those discussed in previous sections, i.e., Eq.~(\ref{eq:constraint}). 
When the bond involves a non-diluted and diluted site with $n_i \neq 0$ and $n_{i+\mu} = 0$, 
then one simply has that $S^\mu_i = B_\mu/K$, so a single component of the remaining spin on the bond is pinned. 

These constraints, if satisfied, have direct implications for the magnetization. Summing over all sites
\begin{equation}
  \sum_i n_i \vec{S}_i = N \frac{\vec{B}}{2K}(1-\rho_2).
\end{equation}
For random dilution, we expect that $\rho_2 = 3\rho^2$ when $\rho \ll 1$ since each site has three neighbours.
Thus, interestingly, diluting the sites does not affect the magnetization if $\rho_2 =  0$. In that case, the magnetization per \emph{spin} is thus increasing, since there are only $(1-\rho)N$ spins left and $\rho \leq 1$. Explicitly,
\begin{equation}
  \vec{M} \equiv \frac{1}{N_s}\sum_i n_i\vec{S}_i = \frac{\vec{B}}{2K}\left(\frac{1-\rho_2}{1-\rho}\right).
  \label{eq:defectmag}
\end{equation}
Alternatively, if we only remove nearest-neighbour pairs, $\rho_2 = 2\rho$ and the magnetization is decreasing.
The susceptibility is also changed similarly, being enhanced by dilution when $\rho_2=0$, with
\begin{equation}
  \chi_{\mu\nu} \equiv \frac{\del M_\mu}{\del B_\nu} = \frac{\delta_{\mu\nu}}{2K}\left(
    \frac{1-\rho_2}{1-\rho}
    \right).
\end{equation}

\begin{figure}[tp]
  \centering
  \includegraphics[width=0.8\columnwidth]{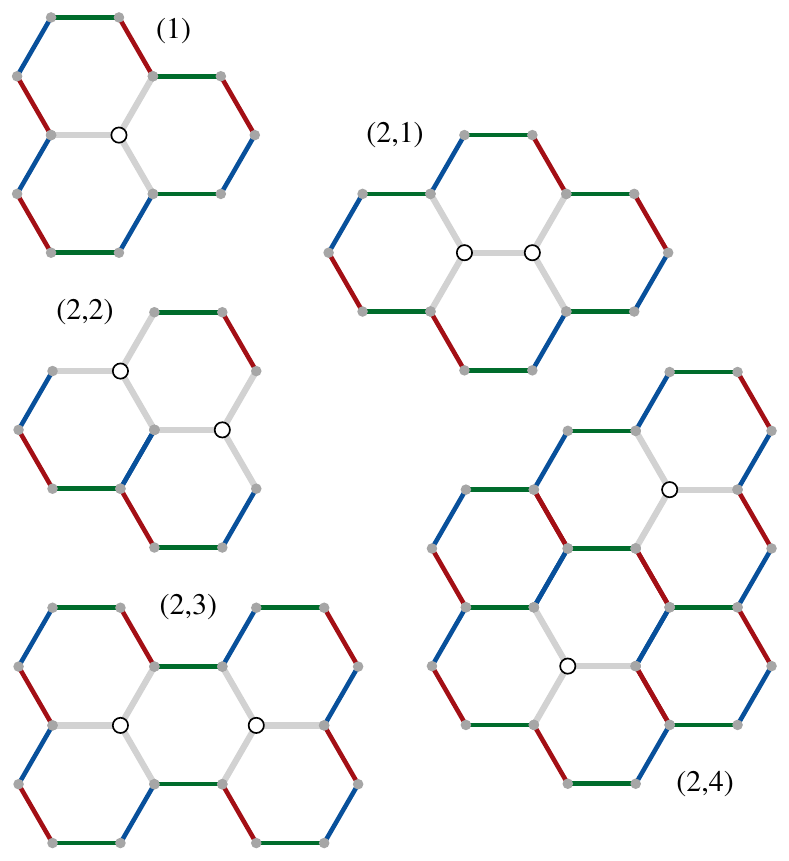}
  \caption{\label{fig:vacancies}
Various one and two vacancy configurations. Each missing spin is associated with three neighbouring spins that each lose a bond. We consider threshold fields for the $[111]$ field direction. For the isolated defect the threshold is $B^{(1)}=\sqrt{3}$. For pairs of vacancies, the threshold fields, $B^{(2,n_b)}$, depend on the minimal number of bonds, $n_b$, between neighboring spins belonging to the two defects. Then for the panels labeled by $(2,b)$ we find $B^{(2,b)}$, defined in the main text, is $B^{(2,1)} = \sqrt{3/2}K$, $B^{(2,2)} =2K\sqrt{3/5}$, $B^{(2,3)} = (K/2)\sqrt{3(2+\sqrt{3})}$, $B^{(2,4)} = 8\sqrt{3/65}K$. 
  }
\end{figure}

In addition to isolated vacancies and dilution on bonds, we can consider some other vacancy configurations as a function of field and the resulting threshold fields. For a single vacancy [Fig.~\ref{fig:vacancies}, configuration (1)] located site $I$, $n_I =0$, the constraints on the spins neighboring the defect $S_{I+\mu}^\mu=B_\mu/K$. Normalization of the spins gives the threshold $B_\mu/K \leq 1$, implying that the threshold field depends on the field orientation. Given $\hat{\vec{B}}$, the threshold is $K/{\rm max}(\hat{B}_x,\hat{B}_y,\hat{B}_z)$ which is $\sqrt{3}K$ for $\vec{B}\parallel [111]$ and $K$ for $\vec{B}\parallel [100]$. Direct numerical minimization of a system with a single defect confirms that the constraints can be satisfied 
up to the predicted threshold field. We see then that presence of defects both reduces the threshold field and introduces a dependence on field direction. We denote this reduced threshold field
as $B^{(1)}_s$, indicating  that it appears due to single defects and thus represents an $O(\rho)$ effect.

For systems with pairs of vacancies, $\rho_2 \neq 0$, the threshold field depends on the precise spatial configuration of these defects. Fig.~\ref{fig:vacancies} shows various configurations of vacancy pairs. 
We illustrate how the threshold fields can be determined for a few of these configurations fixing a $[111]$ field direction. First, consider a pair of defects on neighbouring sites, as illustrated in Fig.~\ref{fig:vacancies}, configuration (2,1). Here the defects are at sites $I$ and $I+z$ and thus we have constraints 
\begin{align}
    S^x_{I+x} =
    S^y_{I+y} = S^x_{I+z+x} = S^y_{I+z+y} = \frac{B}{\sqrt{3}K}.
\end{align}
Each of these constraints requires that $B \leq \sqrt{3}K \equiv  B_s^{(2,1)}$ for the constraints to be satisfiable. We thus see that the nearest neighbour pairs give
the same threshold field as in the single defect case. Next consider a pair of defects that are connected by an undiluted site, as shown in  Fig.~\ref{fig:vacancies}, configuration (2,2).
Label the shared site as $I$ and the diluted sites as $I+x$ and $I+y$.
The constraints for the neighbouring, but unshared sites are of the same character as above, but for the shared site we require
\begin{equation}
S^x_{I} = S^{y}_I = \frac{B}{\sqrt{3}K}.
\end{equation}
This implies that $S^z_I = \pm \sqrt{1-2B^2/(3K^2)}$ which requires that 
\begin{equation}
B  \leq \sqrt{\frac{3}{2}}K 
\end{equation}
We thus see that this configuration reduces the threshold field further to $B^{(2,2)}_s = \sqrt{3/2}K$. 

\begin{figure}[tp]
  \centering
  \includegraphics[width=\columnwidth]{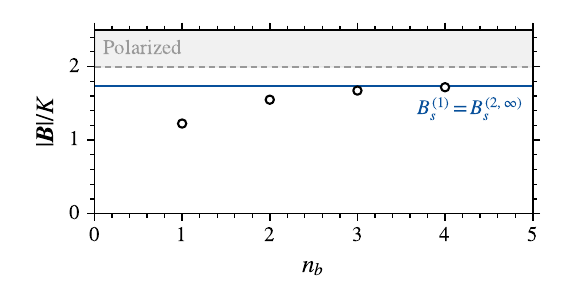}
  \caption{\label{fig:threshold}
Threshold fields, as described in the main text, for pairs of defects $B_s^{(2,n_b)}$ separated by $n_b$ bonds for a $[111]$ magnetic field.
As the separation of defects increases the threshold field rapidly converges to $B^{(1)}_s = B^{(2,\infty)}_s = \sqrt{3}K$.
  }
\end{figure}

We can derive similar inequalities for more complex configurations of defects as shown in Fig.~\ref{fig:vacancies}. The various thresholds in a $[111]$ field are given in the caption and plotted in Fig.~\ref{fig:threshold}. These are computed exactly from the constraint equations and found, empirically, to depend on the minimal number of bonds $n_b$ connecting spins neighboring the vacancies. The threshold $B^{(2,n_b)}_s$ is always bounded as $B^{(2,n_b)}_s\leq\sqrt{3}K$, saturating rapidly as the separation increases. The spin liquid therefore ``heals" very quickly with distance and defects are efficiently screened from one another. 

While the threshold field is always finite for pairs of defects, bounded from below by $B^{(2,1)}_s$, this is not true for larger numbers of defects. For configurations of three defects, there are arrangements that force a threshold field of zero. The relevant configuration is one where a single spin is isolated by removing its neighbors and this isolated spin is then unconstrained. As it is unconstrained, it cannot provide the required compensation of the surrounding defects and the constraints are violated. 

Assembling the ingredients from the foregoing discussion, we can now provide a picture for how the spin liquid behaves as a function of dilution and field. For very small $\rho$, for which $O(\rho^2)$ and higher effects are negligible, the system essentially consists of widely separated isolated defects. Then the magnetization of the spin liquid is identical to that of the $\rho=0$ liquid up to a field orientation dependent threshold with an upper bound of $\sqrt{3}K$ $-$ less than the threshold on the $\rho=0$ liquid of $2K$. The magnetization at threshold is suppressed by a factor of $1-\rho$ which simply accounts for the missing moments. To $O(\rho^2)$ there is a reduction of the threshold field to $\sqrt{3/2}K$ corresponding to a finite density of vacancy configurations of the type shown in Fig.~\ref{fig:vacancies}(b). There is also a net reduction of the magnetization to this order arising from missing bonds [see Eq.~(\ref{eq:defectmag})]. 
On the basis of the isolated spin case we know that, at $O(\rho^3)$, the threshold drops to zero, and the constraints are locally violated, and we expect the effect of these three defect configurations to 
be localized around the defect cluster.

\section{Conclusions and Discussion}

This paper is primarily an addition to the ever-growing, but relatively short, list of classical statistical mechanics models with extensive ground state degeneracy. The first such model to our knowledge is the Pauling model of water ice from the 1930s. Many later models arose in the context of classical frustrated magnetism. Broadly these locally constrained ground states can be classified into models with power law correlations and those with short-range correlations. The ice-like models fall into the former category giving so-called Coulomb phases with local constraints that map to a discrete version of the Gauss law in electrostatics. Also in this class are classical spin liquids with rank two tensor analogues of the electrostatic Gauss law. The classical Kitaev model in zero field is remarkable  as an example of a classical spin liquid with short-range spin correlations and power law correlations in the quadrupoles that is coupled wholly through the dipolar degrees of freedom.~\footnote{A rare example of a phase with pinch points in the quadrupolar correlations that has short-ranged spin correlations is a model with both bilinear and biquadratic couplings \cite{pohle2026spinliquidsspin1pyrochlore}.}

Usually spin liquids are fragile to perturbations so it is remarkable that the application of a natural perturbation -- a magnetic field -- to the classical Kitaev leads to a {\it different} kind of classical spin liquid and over a finite interval of field strengths. We have established the existence of this spin liquid and its short-range correlations that can be viewed as Higgsing out the fluctuating gauge field. We have also shown the perfect screening of the magnetic field in the presence of dilute site disorder. Any future classification of classical spin liquids must incorporate this novel instance. Existing classification schemes that are primarily based on features of the band structure at large $N$ fail to capture the distinguishing features of the Kitaev model at zero field and at finite field. 

No less interesting than the direct phenomenology of the classical Kitaev liquid is the formation of the liquid from the paramagnetic region above the saturation field. As shown previously, the high field regime has a pair of magnon bands with nonzero Chern number. As the saturation field is approached from above, the lowest magnon band flattens and the flat band condenses at saturation. This is a rare instance of finite field flat band condensation and the only instance where that band is topologically nontrivial. It would be interesting to understand better how the flat band feeds through to the properties of the classical spin liquid.  

We have discussed the classical spin liquid of this paper in light of the question of understanding such states of matter. Finally, we remark on the broader interest in Kitaev systems. It is well known that the spin one-half Kitaev model has a zero field spin liquid and a small field chiral spin liquid phase. Various numerical studies have found evidence for an intermediate field regime between the chiral liquid and the trivial polarized state with very soft modes. The nature of this regime has not been settled definitively. This work reveals that an intermediate field regime exists even in the classical case and that it is a liquid. It is an interesting open question to find the state of matter that arises when the classical liquid is subject to weak quantum fluctuations. Does the liquid survive or is it replaced with some kind of long-range order?

\acknowledgements

This work was supported by the Hungarian National Research, Development and Innovation Office (NKFIH) under Grant No. K-142652. PAM acknowledges the CNRS. Work at the University of Windsor (JGR) was funded by the Natural Sciences and Engineering Research Council of Canada (NSERC) (Funding Reference No. RGPIN-2020-04970).
This work was in part supported by the Deutsche Forschungsgemeinschaft under grants SFB 1143 (project-id 247310070) and the cluster of excellence ctd.qmat (EXC 2147, project-id 390858490). 

\appendix

\section{Local zero modes in the canted Néel state}
\label{sec:local_mode}

From Section~\ref{sec:linearizedzeromodes} we recall the constraint equations for deformations, localized on a hexagon, about a canted N\'{e}el state.  
\begin{subequations}
\label{eq:Sieta}
\begin{align}
\eta_{1,6}^2 + 2 S_A^y \eta_{1,6} + \eta_{1,2}^2 + 2 S_A^z \eta_{1,2} &= 0\,, \label{eq:Sieta_a} \\
\eta_{3,2}^2 - 2 S_B^x \eta_{3,2} + \eta_{1,2}^2 - 2 S_B^z \eta_{1,2} &= 0\,, \label{eq:Sieta_b} \\
\eta_{3,2}^2 + 2 S_A^x \eta_{3,2} + \eta_{3,4}^2 + 2 S_A^y \eta_{3,4} &= 0\,, \label{eq:Sieta_c}\\
\eta_{3,4}^2 - 2 S_B^y \eta_{3,4} + \eta_{5,4}^2 - 2 S_B^z \eta_{5,4} &= 0\,, \label{eq:Sieta_d} \\
\eta_{5,6}^2 + 2 S_A^x \eta_{5,6} + \eta_{5,4}^2 + 2 S_A^z \eta_{5,4} &= 0\,, \label{eq:Sieta_e} \\
\eta_{5,6}^2 - 2 S_B^x \eta_{5,6} + \eta_{1,6}^2 - 2 S_B^y \eta_{1,6} &= 0\,. \label{eq:Sieta_f}
\end{align}
\end{subequations}
For the notation for the spins, components and deformations we refer to Fig.~\ref{fig:hexagon}. In Section~\ref{sec:linearizedzeromodes}, we showed that the linearized equations based on Eqs.~\ref{eq:Sieta} have zero modes on the hexagon. Here we go further by explicitly constructing finite deformations for the full nonlinear equations. 

We point out that the system of equations~\eqref{eq:Sieta} is invariant under independent sign flips in each Cartesian sector,
\begin{subequations}
\label{eq:setatrans}
\begin{align}
 (S_A^x,S_B^x,B_x,\eta_{3,2},\eta_{5,6}) &\to (-S_A^x,-S_B^x,-B_x,-\eta_{3,2},-\eta_{5,6})\,, \\
 (S_A^y,S_B^y,B_y,\eta_{3,4},\eta_{1,6}) &\to (-S_A^y,-S_B^y,-B_y,-\eta_{3,4},-\eta_{1,6})\,, \\
 (S_A^z,S_B^z,B_z,\eta_{5,4},\eta_{1,2}) &\to (-S_A^z,-S_B^z,-B_z,-\eta_{5,4},-\eta_{1,2})\,.
\end{align}
\end{subequations}
Although $B_\alpha$ does not appear explicitly in Eq.~\eqref{eq:Sieta}, it is included here because $S_{A}^\alpha + S_{B}^\alpha \propto B_\alpha$.
Consequently, without loss of generality we may restrict to field directions in the octant $B_x\ge 0$, $B_y\ge 0$, and $B_z\ge 0$.

We now derive two relations from Eqs.~\eqref{eq:Sieta}.
First, taking the alternating linear combination
\begin{equation}
\label{eq:altcomb}
(\ref{eq:Sieta_a})-(\ref{eq:Sieta_b})+(\ref{eq:Sieta_c})-(\ref{eq:Sieta_d})+(\ref{eq:Sieta_e})-(\ref{eq:Sieta_f})\,,
\end{equation}
the quadratic terms in the $\eta$'s cancel identically, yielding the linear constraint
\begin{multline}
\label{eq:etalin}
(S_A^x + S_B^x)\,(\eta_{3,2} + \eta_{5,6})
\\
+ (S_A^y + S_B^y)\,(\eta_{1,6} + \eta_{3,4})
\\
+ (S_A^z + S_B^z)\,(\eta_{1,2} + \eta_{5,4})
= 0\,.
\end{multline}

Second, we eliminate the linear terms by forming a weighted sum of Eqs.~\eqref{eq:Sieta_a}--\eqref{eq:Sieta_f}.
Specifically, we multiply
Eq.~\eqref{eq:Sieta_a} by $S_A^x S_B^y S_B^z$,
Eq.~\eqref{eq:Sieta_b} by $S_A^x S_B^y S_A^z$,
Eq.~\eqref{eq:Sieta_c} by $S_B^x S_B^y S_A^z$,
Eq.~\eqref{eq:Sieta_d} by $S_B^x S_A^y S_A^z$,
Eq.~\eqref{eq:Sieta_e} by $S_B^x S_A^y S_B^z$,
and Eq.~\eqref{eq:Sieta_f} by $S_A^x S_A^y S_B^z$.
Upon summation, all terms linear in $\eta_{ij}$ cancel, leaving the quadratic relation
\begin{multline}
\label{eq:etaquad}
(S_A^x+S_B^x)\Bigl(S_A^y S_B^z\,\eta_{5,6}^2+S_B^y S_A^z\,\eta_{3,2}^2\Bigr)
\\
+(S_A^y+S_B^y)\Bigl(S_A^x S_B^z\,\eta_{1,6}^2+S_B^x S_A^z\,\eta_{3,4}^2\Bigr)
\\
+(S_A^z+S_B^z)\Bigl(S_A^x S_B^y\,\eta_{1,2}^2+S_B^x S_A^y\,\eta_{5,4}^2\Bigr)
=0\,.
\end{multline}

\subsection{Analytical solution for a special case}

To assess whether Eqs.~(\ref{eq:Sieta}) admit a continuous family of solutions, we pursued the following strategy. We generated a number of $(\mathbf S_A,\mathbf S_B)$ with rational components (using Pythagorean quadruples) and solved Eqs.~(\ref{eq:Sieta}) numerically to high precision, starting from random initial configurations. Empirically, for generic choices of $\mathbf S_A$ and $\mathbf S_B$ we find only a small discrete set of solutions. However, for certain special Néel states, the procedure consistently returns a "continuum" of solutions (essentially a different solution for each random initial configurations). We take this as a numerical indication of an exact zero-mode manifold, which we prove analytically below.

One representative class of such special backgrounds consists of configurations related by permutations of Cartesian components, for example
\begin{equation}
\label{eq:SASB_uvw}
\vec{S}_A = (u,v,w)\,,\qquad \vec{S}_B = (-u,w,v)\,,
\end{equation}
with $u>0$ and $u^2+v^2+w^2=1$.
For this choice, from Eq.~(\ref{eq:SASBBconstraint}), the corresponding field direction is
\begin{equation}
\label{eq:Bdir_uvw}
  \frac{\vec{B}}{K} = \vec{S}_A + \vec{S}_B = (0,\,v+w,\,v+w)\,,
\end{equation}
which lies on the octant boundary $B_x=0$.
Symmetry-related cases obtained by permuting Cartesian components (e.g. $\vec{S}_A=(v,u,w)$ and $\vec{S}_B=(w,-u,v)$) and by applying the transformations in Eqs.~(\ref{eq:setatrans}) are likewise allowed.

Substituting Eq.~(\ref{eq:SASB_uvw}) into Eqs.~(\ref{eq:etalin}) and~(\ref{eq:etaquad}) yields
\begin{subequations}
\label{eq:etalinq_uv}
\begin{align}
0 &=(v+w)\,(\eta_{1,6}+\eta_{5,4}+\eta_{1,2}+\eta_{3,4})\,, \label{eq:etalinq_uv_a}\\
0 &=u(v+w)\Bigl[v\,(\eta_{1,6}^2-\eta_{5,4}^2)+w\,(\eta_{1,2}^2-\eta_{3,4}^2)\Bigr]\,. \label{eq:etalinq_uv_b}
\end{align}
\end{subequations}
For $u\neq 0$ and $v+w\neq 0$, the generic solution is 
\begin{subequations}
\label{eq:eta_identifications}
\begin{align}
  \eta_{5,4} &= -\eta_{1,6}\,, \\
  \eta_{3,4} &= -\eta_{1,2}\,.
\end{align}
\end{subequations}

With Eq.~(\ref{eq:eta_identifications}), the system \eqref{eq:Sieta} reduces to three coupled equations for the four remaining variables:
\begin{subequations}
\label{eq:Sietauvw}
\begin{align}
-2 w \eta_{1,6}+\eta_{1,6}^2+2 u \eta_{5,6}+\eta_{5,6}^2 &=0\,,\\
2 w \eta_{1,2}+\eta_{1,2}^2+2 v \eta_{1,6}+\eta_{1,6}^2 &=0\,,\\
-2 v \eta_{1,2}+\eta_{1,2}^2+2 u \eta_{3,2}+\eta_{3,2}^2 &=0\,.
\end{align}
\end{subequations}
Thus, we obtain three constraints for four $\eta$ parameters, leaving one continuous degree of freedom. The corresponding spin configuration becomes
\begin{subequations}
\label{eq:Si_uv}
\begin{align}
  \vec{S}_1 & = (u ,\, v + \eta_{1,6},\, w + \eta_{1,2})\,, \\
  \vec{S}_2 & = (-u - \eta_{3,2},\, w,\, v - \eta_{1,2})\,, \\
  \vec{S}_3 & = (u + \eta_{3,2},\, v - \eta_{1,2},\, w)\,, \\
  \vec{S}_4 & = (-u,\, w + \eta_{1,2},\, v + \eta_{1,6})\,, \\
  \vec{S}_5 & = (u + \eta_{5,6},\, v,\, w - \eta_{1,6})\,, \\
  \vec{S}_6 & = (-u - \eta_{5,6},\, w - \eta_{1,6},\, v)\,.
\end{align}
\end{subequations}
Spins are paired as $(1,4)$, $(2,3)$, and $(5,6)$ such that each pair sum is collinear with the field direction $(0,1,1)$, albeit with pair-dependent magnitude:
\begin{subequations}
\label{eq:pair_sums}
\begin{align}
  \vec{S}_1 + \vec{S}_4 & = (v + w + \eta_{1,6} + \eta_{1,2})\,(0 , 1, 1)\,, \\
  \vec{S}_2 + \vec{S}_3 & = (v + w - \eta_{1,2})\,(0, 1, 1)\,, \\
  \vec{S}_5 + \vec{S}_6 & = (v + w - \eta_{1,6})\,(0, 1, 1)\,.
\end{align}
\end{subequations}

We also repeated the numerical procedure for two bond-sharing hexagons. In that geometry, we find numerical evidence for a continuous family of solutions for generic field directions. An analytic treatment, however, appears intractable due to the increased number of coupled constraints.

\subsection{Linear approximation}

In addition to the solution given in Section~\ref{sec:linearizedzeromodes} of the main text, we point out
that by relaxing the assumption of a background canted Néel order, a necessary condition for the existence of a local zero mode is that the components of the spins building the hexagon satisfy
\begin{equation}
S_1^y S_2^z S_3^x S_4^y S_5^z S_6^x = S_1^z S_2^x S_3^y S_4^z S_5^x S_6^y
\end{equation}
It may be worth adding that this plaquette constraint, in the spin one-half case, is a flux constraint on a single plaquette (recalling that the flux quantum number is conserved in the quantum model).

\section{Comments on local symmetries}
\label{sec:local_symmetries}

The zero field ground state constraint is
\begin{equation}
S_i^\mu + S_{i+\mu}^{\mu} = 0
\end{equation}
which is invariant under a local sign change of the components $S_i^\mu$ and $S_{i+\mu}^{\mu}$. So there are three local $\mathbb{Z}_2$ symmetries for each primitive cell. These are sufficient to ensure that correlators $\langle S_i^\mu S_j^\nu \rangle =0$ unless both $i$ and $j$ are neighboring sites and $\mu=\nu$. 

For the finite field case, it is convenient to generalize the model considered in the main text to one where the Zeeman term is replaced by
\begin{equation}
- \sum_{i,\mu} \mathfrak{b}_{i\mu} \left( S_i^\mu + S_{i+\mu}^\mu \right)
\end{equation}
for bond variable $\mathfrak{b}_{i\mu}$. The finite field case corresponds to $\mathfrak{b}_{i\mu} \rightarrow B_{\mu}$. The local constraint is now
\begin{equation}
S_i^\mu + S_{i+\mu}^{\mu} = \frac{\mathfrak{b}_{i\mu}}{K}.
\end{equation}
This model has local $\mathbb{Z}_2$ symmetries $(S_i^\mu,S_{i+\mu}^\mu,\mathfrak{b}_{i\mu})\rightarrow (-S_i^\mu,-S_{i+\mu}^\mu,-\mathfrak{b}_{i\mu})$. If $b_{i\mu}$ are considered to be background fields there is a loose analogy with a Higgs mechanism where $S_{i}^{mu}$ and $S_{i+\mu}^{mu}$ both acquire a vacuum expectation value of $\mathfrak{b}_{i\mu}/2K$. Under the local symmetry, the vacuum configuration changes. In this regime $\langle S_i^\mu S_j^\nu \rangle_{\boldsymbol{\mathfrak{b}}}$ no longer vanishes in general but there are identities connecting correlators for different vacuum configurations. For example $\langle S_i^\mu S_{i+\mu}^\mu \rangle_{\boldsymbol{\mathfrak{b}}}=-\langle S_i^\mu S_{j}^\nu \rangle_{\boldsymbol{\mathfrak{b}}'}$ where, for example, $\mathfrak{b}'_{j\nu}=-\mathfrak{b}_{j\nu}$. It follows that correlators are left unchanged under sign changes of local fields that are not on the bonds associated to the $(i,\mu)$, $(j,\nu)$.

\section{Comments on the Gaussian approximation}

Here we discuss the spin liquids in the classical Kitaev model in relation to recent work categorizing classes of classical spin liquids \cite{davier2023,yan2024,yan2024b,fang2024}. Much insight has been gained over the years by supposing that bilinear exchange couplings are quadratic couplings between variables defined over the real numbers in an otherwise noninteracting model. The spectra of such Gaussian approximations to spin models hosting classical spin liquids tend to contain low-lying flat bands that are interpreted as capturing both the extensive degeneracy of the liquid through a projector onto states satisfying the local constraint. 

The Gaussian theory for the classical Kitaev model in zero field breaks up into three blocks of the form
\begin{equation}
K \left( \begin{array}{cc} 0 & \exp(i\vec{k}\cdot\boldsymbol{\delta}_{\mu}) \\
\exp(-i\vec{k}\cdot\boldsymbol{\delta}_{\mu}) & 0 \end{array} \right)
\end{equation}
for nearest neighbor bonds $\boldsymbol{\delta}_{\mu}$ ($\mu=x,y,z$). The spectrum therefore has three-fold degenerate flat bands at energies $\epsilon=\pm 1$ with eigenvectors
$\trp{(\pm\exp(i\vec{k}\cdot\boldsymbol{\delta}_{\mu}) \ 1 )}$ that wind through the zone in the $\vec{\delta}_{\mu}$ direction.~\footnote{This approach is equivalent to writing the Hamiltonian in the form $\sum_{l\in {\rm cluster}}  |C(l)|^2$ -- such that the energy is minimized when the constraint $C(l)=0$ is satisfied -- and then Fourier transforming the constraints.} Each mode here is essentially one dimensional and the block structure indicates that frustration is absent. As the spectrum is gapped, Gaussian correlations cannot capture the quadrupolar pinch-point correlations in the interacting spin model. Taken together these observations show that the non-linearities inherent to the spin model are fundamental to the physics.

In the presence of a magnetic field, as we have shown, the gapped response can be thought of as originating from a Higgs mechanism thus necessitating fluctuation effects. Another point of view is that the shift in the constant $h^\mu$ in the local constraint $S^\mu_i + S^\mu_{i+\mu} = h^\mu$ away from zero when a magnetic field is introduced does not alter the spectrum in the Gaussian approximation though the physics changes dramatically. 

In summary, the classical Kitaev model both in zero field and at finite field fall outside the simple Gaussian approximation scheme for categorizing classical spin liquids.

\section{Various Field Directions}
\label{app:directions}

The analysis in the main text focused on the $[111]$ field direction, which preserves the $C_3$ symmetry of the lattice. To demonstrate that the field-induced spin liquid is not an artifact of this high-symmetry direction, we present thermodynamic data for magnetic fields applied along the $[100]$, $[110]$, and generic $[11\beta_4]$ directions where $\beta_4 = 2^{1/4}$~\footnote{This direction was chosen from a family of field directions $[11\beta_n]$ where $\beta_n^2 = 2\tan(\pi/(2n))/(1-\tan(\pi/(2n)))$ where it is possible to analytically construct commensurate ground states (with unit cell size of $n$ by $n$ or $2n$ by $2n$) for all field strengths.}.

\begin{figure}[h]
  \centering
  \includegraphics[width=\columnwidth]{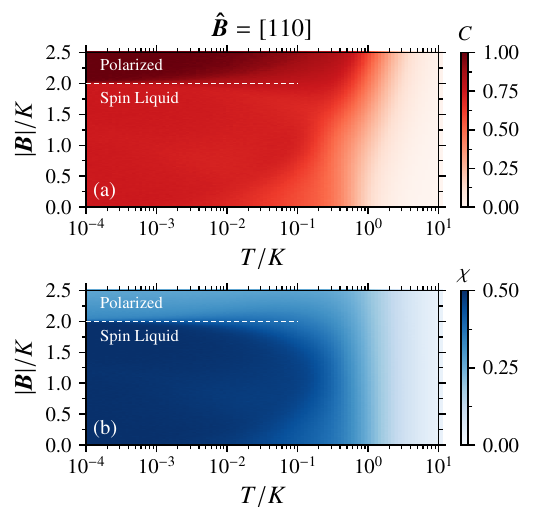}
  \caption{\label{fig:app_110}
  Thermodynamics for a field applied along the $[110]$ direction ($L=8$, $10^5$ sweeps). The heat capacity approaches the expected $3/4$ value at low temperatures, and the susceptibility remains constant within the liquid phase, mirroring the $[111]$ phenomenology.}
\end{figure}

\begin{figure}[h]
  \centering
  \includegraphics[width=\columnwidth]{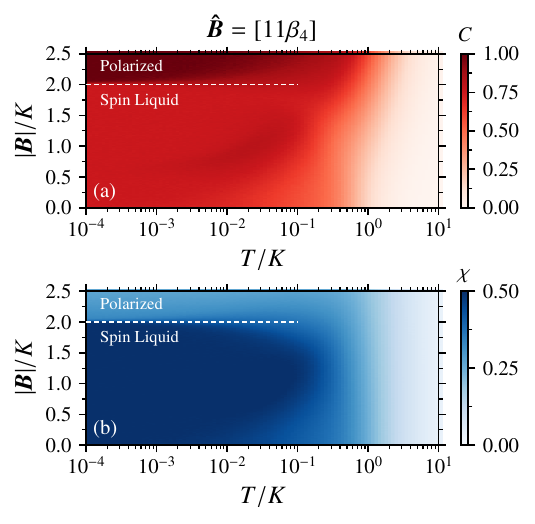}
  \caption{\label{fig:app_11x}
  Thermodynamics for a generic field direction $[11\beta_4]$ where $\beta_4 = 2^{1/4}$ ($L=8$, $10^5$ sweeps). The heat capacity approaches the expected $3/4$ value at low temperatures, and the susceptibility remains constant within the liquid phase, mirroring the $[111]$ phenomenology.}
\end{figure}

Figures~\ref{fig:app_110} and \ref{fig:app_11x} show the heat capacity and magnetic susceptibility obtained from Monte Carlo simulations. In all cases, the behavior is qualitatively identical to the $[111]$ case. The heat capacity shows no sharp features indicative of a phase transition down to the lowest temperatures simulated, confirming the absence of thermal order-by-disorder. Furthermore, the specific heat approaches the characteristic value of $C \to 3/4$ per spin in the low-temperature limit for fields $|\vec{B}| < 2K$, consistent with the counting of zero modes derived in the main text.

Similarly, the magnetic susceptibility remains featureless and isotropic within the spin liquid regime. These results confirm that the macroscopic degeneracy and the associated classical spin liquid phase are robust features of the classical Kitaev model for arbitrary field orientations below the saturation threshold $B_s = 2K$.

\bibliography{draft}

\end{document}